\documentclass[preprint,showpacs,preprintnumbers,amsmath,amssymb]{revtex4}
\usepackage{amsmath}

\usepackage{graphicx}
\usepackage{pdfpages}
\usepackage{dcolumn}
\usepackage{bm}
\usepackage{amsmath} 
\usepackage{amsfonts}
\usepackage{amssymb}
\usepackage{url}
\usepackage{microtype}
\usepackage{graphicx}%

\newcommand{\qed}{\nobreak \ifvmode \relax \else
      \ifdim\lastskip<1.5em \hskip-\lastskip
      \hskip1.5em plus0em minus0.5em \fi \nobreak
      \vrule height0.75em width0.5em depth0.25em\fi}

\begin{document}

\preprint{}
\title{Quantum Steering Ellipsoid Volume as a Measure on the Two-Qubit States and Associated Separability and Absolute Separability Ratios}
\author{Paul B. Slater}
 \email{slater@kitp.ucsb.edu}
\affiliation{%
Kavli Institute for Theoretical Physics, University of California, Santa Barbara, CA 93106-4030\\
}
\date{\today}
            
\begin{abstract}
Employing the volume of quantum steering ellipsoids (QSEs) as a  measure on the fifteen-dimensional convex set of 
two-qubit states, we estimate the  ratio of the integral of the measure over the separable states to its  integral over all (separable and entangled) states to be 0.0288. This can be contrasted with the considerably larger separability ratios (probabilities) of  $\frac{8}{33} = \frac{2^3}{3 \cdot 11} \approx 0.242424$ and $\frac{25}{341}=\frac{5^2}{11 \cdot 31} \approx 0.0733138$ that various forms of evidence point to with the use of the prominent Hilbert-Schmidt and Bures measures, respectively. The questions of whether the ratio in the QSE setting can be more precisely obtained or even exactly 
computed, as well as whether a metric can be constructed, the volume element of which yields the measure, remain to be addressed. We also investigate
related issues pertaining to absolute separability. Further, we examine the behavior of the separability probability--constant in the Hilbert-Schmidt case and decreasing in the Bures--as a function of the Bloch vector norm in the QSE instance. It  appears 
to increase approaching the pure state boundary.
\end{abstract}

\pacs{Valid PACS 03.67.Mn, 02.50.Cw, 02.40.Ft, 02.10.Yn, 03.65.-w}
\keywords{quantum steering ellipsoid, separability probability,  two-qubits,  Hilbert-Schmidt measure, operator monotone functions, Bures measure, steering, ellipsoid, volume, absolute separability, random matrices}

\maketitle
\section{Introduction}
\subsection{Previous two-qubit separability ratio studies}
There exists a variety--actually an infinitude--of possible measures on  quantum states.
Two of particular prominence are the (non-monotone/non-contractive \cite{ozawa2000entanglement,kumar2020wishart}) Hilbert-Schmidt and the (minimal monotone) Bures \cite[secs. 15.3, 15.4]{bengtsson2017geometry}.

In regard to the Hilbert-Schmidt (Euclidean/flat) measure on the fifteen-dimensional set of two-qubit systems, various (both numerical and analytical) forms of evidence strongly indicate--though a formal proof is still lacking--that the ratio (separability probability) of the volume of separable states to the volume of all states (separable and entangled) is $\frac{8}{33} = \frac{2^3}{3 \cdot 11} \approx 0.242424$ \cite{slater2017master,fei2016numerical} \cite[p. 468]{bengtsson2017geometry}. Though clearly not so compelling, a highly extensive numerical analysis points to the separability probability in the Bures two-qubit case  being $\frac{25}{341}=\frac{5^2}{11 \cdot 31} \approx 0.0733138$ \cite[sec. II.C.1]{slater2019numerical}. (The Hilbert-Schmidt  qubit-{\it qutrit} separability probability has been conjectured to equal
$\frac{27}{1000} =\frac{3^3}{2^3 \cdot 5^3}= 0.027$ \cite[sec. III.A]{slater2019numerical} \cite[Tab. 1]{khvedelidze2018generation}. \cite[eq.(33)]{milz2014volumes}. A formula--$1-\frac{256}{27 \pi^2} \approx 0.0393251$--for the two-qubit separability probability associated with the operator monotone [geometric-mean] function $\sqrt{x}$ was advanced in \cite[eq. (87)]{slater2017master}, but since the corresponding volume of states is infinite, not directly supportable in related numerical analyses   \cite{slater2020quasirandom}.) 

The body of research just indicated  largely stems from the much-cited 1998 paper, "On the volume of the set of 
separable states" \cite{zyczkowski1998volume}.
\subsubsection{Absolute separability analyses}
The Hilbert-Schmidt measure has further been applied in the context of the two-qubit states, as regards the absolutely separable states \cite{halder2019characterization}--those that can not be entangled through global unitary transformations. 
We note that the criterion for absolute separability of two-qubit states is that the inequality \cite{verstraete2001maximally,hildebrand2007positive} \cite[eq. (16.79)]{bengtsson2017geometry},
\begin{equation} \label{AbsSepTest}
 \lambda_1-\lambda_3 < 2 \sqrt{\lambda_2 \lambda_4},   
\end{equation}
is satisfied, where $\lambda_1 \geq \lambda_2 \geq \lambda_3 \geq \lambda_4$ are the decreasingly-ordered density-matrix eigenvalues--the four nonnegative eigenvalues summing to 1, by the unit trace requirement.

A highly challenging (Mathematica) computation \cite{JimB} \cite[eq. (34)]{slater2009eigenvalues} yielded that the 
two-qubit Hilbert-Schmidt absolute separability probability is
\begin{equation} \label{JimBresult}
 \frac{29902415923}{497664}-\frac{50274109}{512 \sqrt{2}}-\frac{3072529845 \pi }{32768
   \sqrt{2}}+\frac{1024176615 \cos ^{-1}\left(\frac{1}{3}\right)}{4096 \sqrt{2}} \approx    0.00365826.
\end{equation}
Further, recent computations demonstrate that the Hilbert-Schmidt volume of the maximal ball contained within the absolutely separable states \cite{adhikari2020constructing,Balls} is the lesser amount,
\begin{equation}
\frac{35 \pi }{23328 \sqrt{3}} \approx 0.00272132,
\end{equation}
while that of the minimal ball containing the absolutely separable two-qubit states is the larger,
\begin{equation}
\frac{35 \sqrt{\frac{1}{3} \left(2692167889921345-919847607929856 \sqrt{6}\right)} \pi}{27518828544} \approx 0.0483353.    
\end{equation}
Rather than the 15-dimensional convex set of two-qubit states, restricting attention to the 3-dimensional convex set of ordered spectra of absolutely separable states, the (flat measure) counterparts of the previous three results, following the same  ordering, are \cite{Balls}
\begin{equation}
 \frac{1}{576} \left(8-6 \sqrt{2}-9 \sqrt{2} \pi +24 \sqrt{2} \cos^{-1}\left(\frac{1}{3}\right)\right) \approx 0.00227243,   
\end{equation}
\begin{equation}
 \frac{\pi }{864 \sqrt{3}} \approx 0.0020993   
\end{equation}
and
\begin{equation}
\frac{\left(14-3 \sqrt{6}\right) \pi }{3456 \sqrt{3}} \approx 0.0034909.    
\end{equation}
\subsection{Present study}
Another possible measure on the two-qubit states--although not apparently to this point in time, having been analyzed as such--is the volume of quantum steering ellipsoids (QSEs) \cite{jevtic2014quantum,mccloskey2017einstein} \cite[sec. V.B]{uola2020quantum}--a measure of quantum correlations. (The question of whether a metric $g_{ij}$ can be constructed, the volume element of which, $\sqrt{g}$, yields the QSE measure, merits investigation, as well as the possible monotonicity 
of the metric \cite[eq. (15.2)]{bengtsson2017geometry}.)

The expression for the volume of the ellipsoid $\mathcal{E}_A$--the set of states to which Bob can steer Alice, forming an ellipsoid in Alice's Bloch sphere, considering all possible measurements on his qubit--provides a compact and non-trivial relation between the steering properties of the 
two-qubit $4 \times 4$ density matrix $\rho$ and the ranks of $\rho$ and the partial transpose with respect to Bob's qubit $\rho^{T_B}$ (obtained by transposing in place the four $2 \times 2$ blocks of $\rho$ \cite[eq. (16.52)]{bengtsson2017geometry}).
The volume $V$ of any ellipsoid is proportional to the product of its three semiaxes $V=\frac{4\pi}{3}s_1s_2s_3$. Therefore $\mathcal{E}_A$ has--adopting the notation/argument of \cite{jevtic2014quantum}--volume $V_A = \frac{4\pi}{3}\sqrt{\det Q_A}$, using the ellipsoid matrix \cite[eq. (2)]{jevtic2014quantum}. This may be rewritten as $V_A=\frac{4\pi}{3}\frac{\left\vert\mathrm{det}\Theta\right\vert}{(1-b^2)^2}$--with $\Theta=
\begin{pmatrix}
    1 & \boldsymbol{b}^T \\
    \boldsymbol{a} & T
\end{pmatrix}$, where $\boldsymbol{a},\boldsymbol{b}$ are the Bloch vectors of the reduced states $\rho_A$ and $\rho_B$ of $\rho$ respectively, and $T$ is a 3 $\times$ 3 matrix encoding the correlations, being the correlation matrix. But since $\mathrm{det} \Theta=16 (\mathrm{det} \rho^{T_B} - \mathrm{det} \rho)$, it follows that 
($b$ is the norm of Bob's Bloch vector)
\begin{equation}
\label{VolumeInRho}
V_A = \frac{64\pi }{3}\frac{\left\vert \det  \rho
-\det  \rho^{T_B}  \right\vert }{\left( 1-b^{2}\right) ^{2}}.
\end{equation}

The  volume $V_B$ of $\mathcal{E}_B$ (the set of states to which Alice can steer Bob, forming an ellipsoid in Bob's Bloch vector) can be computed from $V_A$ via the  relation
$V_B = \frac{(1-b^2)^2}{(1-a^2)^2}V_A$. We tested our algorithm by confirming that for the Werner state on the separable-entangled boundary, it yielded the maximal sphere volume  $V_{*}=\frac{4 \pi}{81}$ inscribed inside the largest possible tetrahedron that can be inscribed inside the unit sphere. Any state with a larger volume than  $V_{*}$ must be entangled.

Let us note with regard to the terms in (\ref{VolumeInRho}) that in \cite{slater2015generalized} it was concluded that the apparent Hilbert-Schmidt two-qubit separability probability of $\frac{8}{33}$ is {\it equally} divided (cf. \cite{szarek2006structure}) between those separable density matrices for which $\det  \rho 
>\det  \rho^{T_B}  $ and those for which $\det  \rho
<\det  \rho^{T_B}  $. It was also noted in \cite[sec. 2.2]{slater2015generalized} that $\det  \rho ^{T_B}
>\det  \rho  $ ranges over $[-\frac{1}{16},\frac{1}{432}]$, with the subrange $[0,\frac{1}{432}]$ containing only separable 
states. 

Further, in  \cite[eq. (7)]{slater2015generalized}, a formula 

\[
F_{2}\left(  n,k\right)  =\left\langle \left\vert \rho\right\vert ^{k}\left(
\left\vert \rho^{PT}\right\vert -\left\vert \rho\right\vert \right)
^{n}\right\rangle /\left\langle \left\vert \rho\right\vert ^{k}\right\rangle
\] %
was reported for the moments ($PT$ denoting partial transpose).
It was further expressible as 
\begin{equation} \label{momentformula}
 F_{2}\left(  n,k\right)  =g\left(  k,n\right)  F_{2}^{\prime}\left(
n,k\right)  ,
\end{equation}
where
\[
g\left(  k,n\right)  :=\frac{\left(  k+1\right)  _{n}\left(  k+1+\alpha
\right)  _{n}\left(  k+1+2\alpha\right)  _{n}}{2^{6n}\left(  k+3\alpha
+\frac{3}{2}\right)  _{n}\left(  2k+6\alpha+\frac{5}{2}\right)  _{2n}},
\]%
(embracing the two-re[al]bit [$\alpha=1$], two-qubit [$\alpha=2]$ and two-quater[nionic]bit cases [$\alpha=4$], $\alpha$ denoting the "Dyson-index" of random-matrix theory \cite{slater2008extended})
and 
\begin{align} 
F_{2}^{\prime}\left(  n,k\right)   &  =\frac{\left(  -1\right)  ^{n}\left(
\alpha\right)  _{n}\left(  \alpha+\frac{1}{2}\right)  _{n}\left(
-2k-2n-1-5\alpha\right)  _{n}}{\left(  -n-k\right)  _{n}\left(
-k-n-\alpha\right)  _{n}\left(  -k-n-2\alpha\right)  _{n}}\label{F2p}\\
&  \times\sum_{j=0}^{[n/2]}\frac{\left(  -\frac{n}{2}\right)  _{j}\left(  \frac
{1-n}{2}\right)  _{j}\left(  k+1+\alpha\right)  _{j}\left(  k+1+2\alpha
\right)  _{j}}{j!\left(  1-n-\alpha\right)  _{j}\left(  \frac{1}{2}%
-n-\alpha\right)  _{j}\left(  n+2k+2+5\alpha\right)  _{j}}2^{2j+2n-2j}%
\nonumber\\
&  =\left(  -1\right)  ^{n}2^{2n}\frac{\left(  \alpha\right)  _{n}\left(
\alpha+\frac{1}{2}\right)  _{n}\left(  n+2k+2+5\alpha\right)  _{n}}{\left(
k+1\right)  _{n}\left(  k+1+\alpha\right)  _{n}\left(  k+1+2\alpha\right)
_{n}}\nonumber\\
&  \times~_{4}F_{3}\left(
\genfrac{}{}{0pt}{}{-\frac{n}{2},\frac{1-n}{2},k+1+\alpha,k+1+2\alpha
}{1-n-\alpha,\frac{1}{2}-n-\alpha,n+2k+2+5\alpha}%
;1\right)  .\nonumber
\end{align}
\section{Analysis}
We chose--for our initial, exploratory purposes--a relatively simple-to-implement approach to estimating the average (separable and total) QSE valumes over the 15-dimensional space. At each iteration of the procedure, seeking to generate feasible two-qubit density matrices, we obtained four potential diagonal entries, by first obtaining three values uniformly randomly distributed over [0,1]. Then, we sorted the three values in increasing order, and appended 0 to the beginning, and 1 to the end of the three-fold sequence. The candidate four diagonal entries were then chosen as the differences between successive entries of the five-fold sequence \cite{smith2004sampling} (cf. \cite[App. A]{zyczkowski1998volume}).

Further, the six potential off-diagonal real entries and the coefficients of the six off-diagonal imaginary entries were all chosen as uniform random values in the interval [$-\frac{4}{15},\frac{4}{15}$]. (The interval [-1,1] would be ideal, but the probability of then generating positive semidefinite
$4 \times 4$ Hermitian matrices would be too small for us to obtain estimates in acceptable amounts of time. So, we compromised with the choice of $\frac{4}{15}$.) Roughly speaking, every ten thousand iterations yielded a single feasible density matrix for further analysis. The use of the value $\frac{4}{15}$, along with a flat measure on the states, led to a separability ratio  estimate of 0.246423, quite close to the apparently valid Hilbert-Schmidt two-qubit volume ratio of $\frac{8}{33} \approx 
0.242424$ \cite{slater2017master}. 

Alternatively, pursuing our main objective here, use of the ellipsoid volume $V_A$ (\ref{VolumeInRho}) as the measure on the two-qubit states gave us an estimated QSE separability ratio of 
0.0288754.

To be more specific, in six hundred million iterations, we obtained 58,497 feasible density matrices, of which 14,415 had positive partial transposes (with $\frac{14415}{58497} \approx 0.246423$). The average of the ellipsoid volumes obtained was 0.197412, of which 0.00570036 was contributed by separable states, yielding the just reported QSE separability ratio of 0.0288754.
(We also--from an initial abundance of caution--performed the computations replacing $b^2$ by $a^2$--the norm-squared of Alice's Bloch vector--in the ellipsoid volume formula (\ref{VolumeInRho}), and obtained the quite close value of 0.0289646.)

Out of general interest and relevance here, let us excerpt the following passage from the caption to Fig. 2 in \cite{jevtic2014quantum}: "For separable states, the set $\mathcal{E}_A$ can also be 2-dimensional (a steering pancake), or 1-dimensional (a steering needle), or trivially 0-dimensional. For these cases, steering is either ``complete" if all ensemble decompositions of $\mathbf{a}$ in $\mathcal{E}_A$ are attainable (when the span of $\mathcal{E}_B$ contains $\frac{1}{2} \mathbb{I}$), otherwise the steering is `incomplete". 

Of course, such indicated special cases have zero three-dimensional 
volume--not contributing to our estimate. Let us also observe that steering ellipsoids in a tetrahedron in the Bloch sphere correspond to separable states.

In a second analysis, also employing six hundred million iterations, we obtained 58,959 feasible density matrices,  14,520 of them having positive partial transposes, of which only 209 passed the test (\ref{AbsSepTest}) for absolute separability. 
(We had not implemented this test in the initial analysis.) This led to  an estimate of 0.000352681 for the two-qubit absolutely separable probability based on the 
use--once again--of QSE volume  as the measure.
In the Hilbert-Schmidt case, the ratio of the absolute separability probability (\ref{JimBresult}) to the apparent (total) separability probability $\frac{8}{33}$ is considerably larger, that is 
0.0150903.

Pooling these two six hundred million iterations analyses, leads to a Hilbert-Schmidt estimate of 0.246348, and a geometric steering ellipsoid estimate of 0.0288312--our best current estimate.

To further explore the behavior of our main computational procedure (in which we have restricted off-diagonal real and complex entries to lie in $[-\frac{4}{15},\frac{4}{15}]$, rather than the ideal, but computationally infeasible $[-1,1]$), employing 460 million iterations (generating 45,444 feasible density matrices), we estimated the Bures separability probability \cite[eq. (15.48)]{bengtsson2017geometry} to be 0.0818048. (The Hilbert-Schmidt estimate obtained here was 0.245709, again close to $\frac{8}{33} \approx 0.242424$. The Bures measure is more localized on states of high purity than the Hilbert-Schmidt 
\cite[p. 420]{bengtsson2017geometry}.)

Previous extensive numerical analyses have suggested a Bures two-qubit separability probability conjecture of the lesser value, $\frac{25}{341}=\frac{5^2}{11 \cdot 31} \approx 0.0733138$.
\cite[sec. II.C.1]{slater2019numerical}.  The weight we applied in the Bures analysis here to each feasible density matrix ($\rho$) generated was ($N=4$, with $\lambda$'s being the eigenvalues of $\rho$),
\begin{equation}
  \frac{\prod _{j=1}^{N-1} \prod _{k=j+1}^N \frac{1}{\lambda _j+\lambda _k}}{\sqrt{\prod
   _{i=1}^N \lambda _i}} . 
\end{equation}
This is proportional to the ratio of the volume differentials, $\frac{\mbox{d}V_{Bures}}{\mbox{d}V_{Hilbert-Schmidt}}$
\cite[eqs. (15.47), (15.34)]{bengtsson2017geometry}--since we anticipated and found  that the feasible density matrices generated by our iterative procedure are themselves distributed (roughly--due to the $\frac{4}{15}$ cutoff) according to Hilbert-Schmidt 
measure. (We also conducted earlier exploratory analyses using $\mbox{d}V_{Bures}$--without division by $\mbox{d}V_{Hilbert-Schmidt}$--as the weight. In 380 million iterations [yielding a Hilbert-Schmidt result of 0.243599] this gave us a separability probability estimate of 0.0582719, somewhat more remote to the conjectured value of $\frac{25}{341}=\frac{5^2}{11 \cdot 31} \approx 0.0733138$ than 0.0818048. This latter over-estimation might be suggestive that the central/main 0.0288 QSE estimate is also somewhat high.)

\begin{figure}
    \centering
    \includegraphics{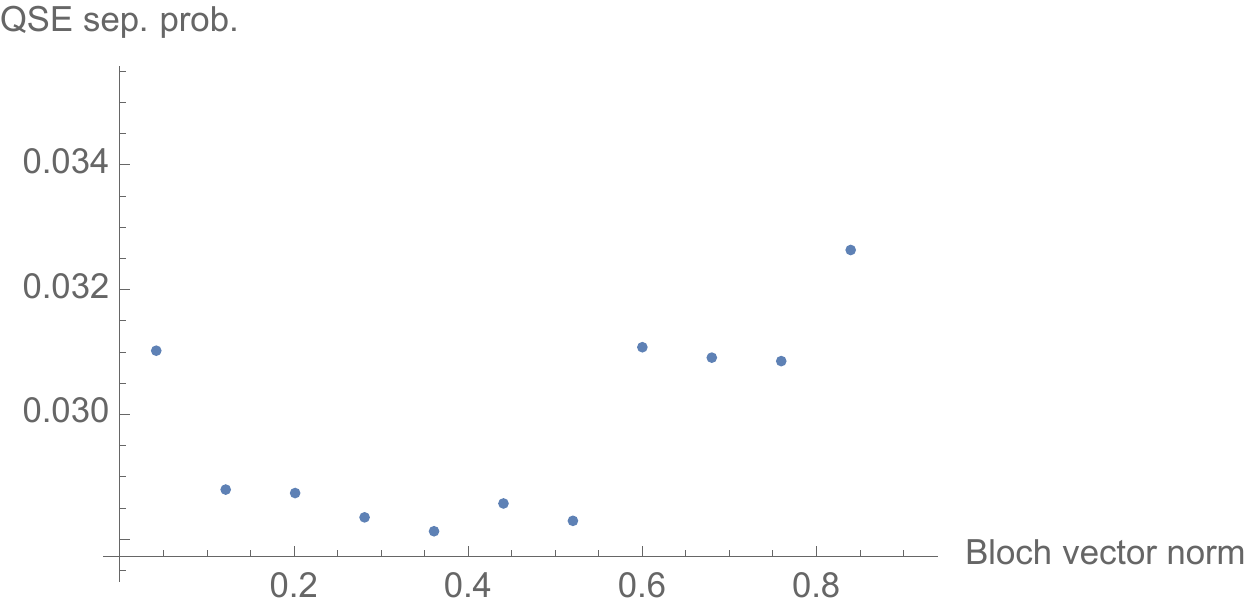}
    \caption{QSE separability probability of Alice's Bloch sphere as a function of the norm $b$ of Bob's Bloch vector.}
    \label{fig:BlochNorm}
\end{figure}

We were further interested in seeing how in the QSE case, the separability probability varies--being constant in the Hilbert-Schmidt instance \cite{milz2014volumes,slater2016invariance} \cite[Cor. 2]{lovas2017invariance} (jumping to 1 at the pure state boundary) and decreasing in the Bures case \cite[Fig. 31]{slater2015bloch}--as a function of the Bloch vector norm (cf. \cite{slater2016invariance}). An analysis (Fig.~\ref{fig:BlochNorm}), employing  one billion iterations of our procedure, generating 97,904 feasible density matrices, 24,304 being PPT, appears to show a degree of increasing behavior as the pure state boundary is approached. (The estimated Hilbert-Schmidt separability probability in this additional analysis was 0.245506.)
\section{Concluding Remarks}
Figure 1 in the somewhat related study \cite{mccloskey2017einstein} shows the distribution of the normalized volume of QSEs versus 
linear entropy--defined as $\frac{4}{3} (1-\mbox{Tr}[\rho^2])$--for $10^8$ random two-qubit states. (Also, Figs. 2 and 4 use 
such states.) The methodology for generating the states is not identified. The authors there seek to  identify  families of states that maximize certain quantifiers of EPR-steering and the volume of the QSE. (Above the linear entropy value of 0.76, they find no states with "positive steerability"  for the measures of "steerable weight" and "robustness of steering".) The focus 
here, contrastingly, has been on the "families" of separable and absolutely separable states.

Let us note that, certainly, more sophisticated estimation procedures can be employed than
the quick, simple-to-implement iterative approximate one we have pursued here in this exploratory analysis. In particular, use of the Euler-angle parameterization of the two-qubit states would seem appropriate \cite{tilma2002parametrization,tilma2002generalized,slater2005qubit}. (Random-matrix-based methodologies have been developed 
for the generation of density matrices, randomly distributed with respect to the Hilbert-Schmidt, Bures and induced measures \cite{al2010random} \cite[pp. 421-422]{bengtsson2017geometry}, but apparently not for other measures of potential interest.)

A challenging task would appear to be the adaptation of the moment formula (\ref{momentformula}) \[
F_{2}\left(  n,k\right)  =\left\langle \left\vert \rho\right\vert ^{k}\left(
\left\vert \rho^{PT}\right\vert -\left\vert \rho\right\vert \right)
^{n}\right\rangle /\left\langle \left\vert \rho\right\vert ^{k}\right\rangle
\]
to the specific QSE case ($k=0$, Dyson-index $\alpha=2$), but with the introduction
of the absolute value function and denominator of $(1-b^2)^2$, as in the formula (\ref{VolumeInRho}) for $V_A$.
With such a moment formula, probability-distribution reconstruction methods \cite{provost2005moment} could be employed, as they were in \cite{slater2015formulas,slater2015generalized}, for high-accuracy estimation of the QSE separability ratio.

A two-qubit state is separable--and can be written as the convex sum of at most four products of pure states--if and only if its ellipsoid fits inside a tetrahedron that itself fits inside the Bloch sphere \cite{milne2015euler}.
So, one can pose the surely formidable problem of estimating/determining the ratio (certainly  greater than 0.0288) of the volume of nested tetrahedra to the volume of all ellpsoids (whether for separable states or not).

\begin{acknowledgements}
This research was supported by the National Science Foundation under Grant No. NSF PHY-1748958.
\end{acknowledgements}

\bibliography{main}

\end{document}